\documentclass[%
 reprint,
superscriptaddress,
nofootinbib,
 amsmath,amssymb,
 aps,prd,
]{revtex4-2}
\usepackage{amsmath}
\usepackage[section]{placeins}
\usepackage[utf8]{inputenc}
\usepackage[english]{babel}
\usepackage{float}
\usepackage[dvipsnames]{xcolor}
\usepackage{soul}
\usepackage{graphicx}
\usepackage{dcolumn}
\usepackage{bm}

\usepackage{graphicx}
\usepackage{dcolumn}
\usepackage{bm}

\usepackage[draft, inline, nomargin]{fixme}
\fxsetup{theme=color, mode=multiuser}

\FXRegisterAuthor{bc}{1}{\color{red}BC}
\FXRegisterAuthor{ag}{2}{\color{blue}AG}
\FXRegisterAuthor{ph}{3}{\color{magenta}PH}

\begin{document}


\title{Passive low-energy nuclear recoil detection with color centers}

\author{Bernadette K. Cogswell}
\email{bkcogswell@vt.edu}
\affiliation{Center for Neutrino Physics, Physics Department, Virginia Tech, Blacksburg, VA 24061}
\author{Apurva Goel}

\email{apurva1@uw.edu}
\affiliation{Department of Physics and Department of Astronomy, University of Washington, Seattle 98105}

\author{Patrick Huber}
\email{pahuber@vt.edu}
\affiliation{Center for Neutrino Physics, Physics Department, Virginia Tech, Blacksburg, VA 24061}

\date{\today}%

\begin{abstract}
Crystal damage events such as tracks and point defects have been used to record and detect radiation for a long time and recently they have been proposed as a means for dark matter detection.  Color centers can be read out optically and we propose a scheme based on selective plane illumination microscopy for micrometer-scale imaging of large volumes corresponding to kilogram mass detectors. This class of detectors would be passive and would operate at room temperature and we call this the PAssive Low Energy Optical Color CEnter Nuclear rEcoil (PALEOCCeNe) detection method. We apply these concepts to the detection of reactor neutrinos using coherent elastic neutrino nucleus scattering (CEvNS). Crystal damage  formation energies are intrinsically on the order of $25\,$eV,  resulting in similarly low nuclear recoil thresholds. This would enable the first observation of reactor neutrino CEvNS with detectors as small as 10\,g. Additionally, a competitive search for spin-dependent dark matter scattering down to a dark matter mass of $0.3\,$GeV could be possible. Passive crystal detectors might also be attractive for nuclear non-proliferation safeguards if used to monitor reactor power and to put limits on plutonium production. The passive nature and small footprint of the proposed detectors implies that these might  
fit well within accepted reactor safeguards operations.

\end{abstract}

\maketitle

\section{\label{sec:level1}Introduction}

Detection of low-energy nuclear recoil events is one of the frontiers of experimental particle physics. It was first studied by Drukier and Stodolsky~\cite{drukier} looking for a method to detect coherent elastic neutrino nucleus scattering (CEvNS). This in turn inspired the idea to look for dark matter by direct detection of nuclear recoils~\cite{Goodman:1984dc}.  The present push for nuclear recoil thresholds of less than 1,000\,eV is due to our current understanding of dark matter, which has evolved from a classical electroweak-scale WIMP with $m_\chi\sim 100\,$GeV to a much broader class of candidates, many with mass 1\,GeV or smaller ~\cite{Battaglieri:2017aum}. This, in turn, puts a premium on the smallest possible nuclear recoil thresholds. At the same time, CEvNS, which was first postulated in the 1970s by Freedman~\cite{Freedman:1973yd}, has been discovered in 2017~\cite{Akimov:2017ade} using 50\,MeV neutrinos. A pivotal next step is to demonstrate a measurement of CEvNS with reactor neutrinos of around 4\,MeV in neutrino energy. On this front a large number of CEvNS experiments are under way: CONUS~\cite{Bonet:2020awv}, MINER~\cite{Agnolet:2016zir}, 
NUCLEUS~\cite{Rothe:2019aii}, RED-100~\cite{Akimov:2019ogx},  NEON~\cite{Choi:2020gkm}, RICOCHET~\cite{Billard:2016giu}, TEXONO~\cite{Wong:2016lmb}, CONNIE~\cite{Aguilar-Arevalo:2019jlr}. This strong interest in reactor-based CEvNS detection arises from both basic physics~\cite{deGouvea:2020hfl,Fernandez-Moroni:2020yyl,Miranda:2020zji,Miranda:2020syh,Tomalak:2020zfh} and potential applications to reactor monitoring for nuclear security~\cite{Cogswell:2016aog,Bernstein:2019hix,Bowen:2020unj}.

The idea of using crystal damage as a detection modality for nuclear recoils was developed for dark matter searches~\cite{Essig:2016crl} and subsequently explored both for damage tracks~\cite{Drukier_2019,Edwards_2019} and point defects~\cite{Budnik:2017sbu,Rajendran2017}. The results presented here are an adaptation of these prior findings to reactor neutrino CEvNS and a refinement of some of the prior calculations. Track-based detection faces severe challenges, due to the very low energy of nuclear recoils from reactor neutrinos, and so we focus on point defects, specifically  color centers. Color centers have attracted attention in quantum information science~\cite{Doherty_2013} and metrology~\cite{nv}, since they can be individually detected by fluorescence spectroscopy.
We introduce an optical, non-destructive readout scheme based on so-called light sheet microscopy, also known as selective plane illumination microscopy (SPIM)~\cite{lightsheet}, which is scalable to kilogram quantities of detector material. 
The resulting detectors are passive, room-temperature, sub-kilogram systems with applications to direct dark matter searches, CEvNS detection of reactor neutrinos and reactor monitoring for nuclear non-proliferation and treaty verification.

In any experiment aiming at low-energy recoil detection, be it for dark matter searches or reactor CEvNS, backgrounds are what sets the scale for sensitivity and the required effort. Our proposal is no different in this context. We would like to stress, that nuclear recoil backgrounds at nuclear recoil energies of less than a few hundred electronvolt, the range we are considering here, are essentially a {\it terra incognita} and the focus of intense experimental efforts. Many experiments do see a sharp increase in their background rate towards this region in energy and this trend seems to apply to many different detector technologies, materials and shielding configurations~\cite{excess}. If the cause if this increase is indeed  due to genuine nuclear recoils caused by neutrons then CEvNS detection at reactors may turn out to be very difficult or even impossible, entirely \emph{independent} of the employed detection technology. We will therefore show all our results as a function of the background rate.

In Sec.~\ref{sec:damage} we discuss the formation mechanisms for permanent crystal damage, describe our simulations and introduce the general features for nuclear recoil detection. We also present a readout scheme and derive criteria for material selection. In Sec.~\ref{sec:results} the results for direct dark matter searches, CEvNS detection at reactors and reactor monitoring are presented. We conclude with a summary and outlook in Sec.~\ref{sec:summary}. In the Appendix we provide a detailed calculation of the required parameters.

\section{Crystal damage}
\label{sec:damage}

A recoiling nucleus eventually will loose all of its kinetic energy to the host crystal lattice via a combination of primarily: 
\begin{enumerate}
    \item Phonons -- A collision between the ion and one of the crystal atoms will result in the creation of phonons, which then quickly dissipate. Hence, they are not suitable for the discussion here, except in that any energy deposited in phonons is not available for ionization of lattice defect creation.
   
    \item Ionization -- The interaction of the  electric field of the moving ion and the electrons in the crystal can displace electrons from their lattice sites, creating a net surplus of positive charge. In an insulator, electrons cannot move to fill in these regions of excess positive charge. The precise mechanism by which ionization creates a permanent track that can, for instance, be made visible by etching is not entirely clear. One possible explanation is that the interaction of these positive charges with themselves and with the rest of the crystal induces mechanical strain. It is thought that it is this strain which weakens chemical bonds~\cite{ILIC2003179}.
  
    \item Lattice defect creation -- In cases where the energy transfer to a crystal atom exceeds the binding energy in the crystal, the atom can be dislodged from its lattice site and start to move through the crystal. This can result in a vacancy in the crystal lattice. If sufficient kinetic energy remains in the recoiling atom, it can create a secondary damage cascade, inducing further vacancies.
\end{enumerate}

The use of lattice defects for dark matter detection has been explored by several authors~\cite{Budnik:2017sbu,Rajendran2017}. In addition, the use of ionization tracks has been investigated in a series of papers~\cite{Drukier_2019,Edwards_2019}.  More recently, ionization track detectors have been proposed to study supernova neutrinos~\cite{Baum:2019fqm}, the time evolution of solar~\cite{Arellano:2021jul} and atmospheric~\cite{Jordan:2020gxx} neutrino fluxes. Here, we extend these aforementioned results by performing detailed simulations using the TRIM package~\cite{TRIM} and specifically apply these results towards CEvNS with reactor neutrinos. 

Track-based detection has been studied in considerable detail~\cite{Drukier_2019,Edwards_2019}; thus, we refer the reader to these references for the underlying formation mechanism.  Also, as we discuss in Sec.~\ref{sec:general}, for recoil energies below a few 100\,eV, track detection hits practical limitations since the resulting tracks are very short ($<5\,$nm). 

Nuclear recoils can lead to the displacement of an atom from its lattice site. This site  can remain unoccupied, leading to a vacancy, or it can be filled either by the initial projectile or by some other dislocated atom, in the case of a larger recoil cascade. These are the primary mechanisms for radiation damage from neutron irradiation and, thus, are highly relevant for nuclear engineering. Neutron damage to the reactor vessel is ultimately the limiting factor for the lifetime of a nuclear fission reactor and for future fusion reactors as well. In our exposition, we follow the review in Ref.~\cite{Nordlund:2018}. Frenkel defects, where an atom leaves its lattice site and becomes interstitial, occur naturally in crystals at non-zero temperatures. The lifetime of the these defects is short, since lattice strain associated with the existence of the Frenkel defect will lead to spontaneous recombination.

A permanent vacancy forms when the atom is removed far enough from its lattice site so that recombination is no longer possible. This distance is on the order of 1--2\,nm. Since the stopping power for most ions is around 20--100\,eV/nm, an approximate energy in the range 20--200\,eV is required for the creation of a permanent vacancy. This energy is, at best, indirectly related to the binding energy of a given solid or the defect formation energy.

It makes sense to introduce the concept of a threshold damage energy (TDE), with the idea that below the TDE no permanent vacancies are formed. In reality, the situation is more complicated since the energy required is not isotropic. Also, thermal effects cannot be entirely neglected. Within the simple model of a TDE, denoted as $E_d$, the number of vacancies $n_v$ formed would be 0 for $T_d<E_d$, 1 for $E_d\leq T_d<2 E_d$ and so forth. Note, that $T_d$ is usually called the damage energy, i.e., the kinetic energy remaining in the projectile after all ionization losses have been accounted for.  Kinchin and Pease demonstrated in 1955~\cite{Kinchin} that this would be too simplistic and a better approximation is given by $n_v=T_d/(2 E_d)$ for $T_d\gg E_d$. This can be intuitively understood by the fact that, for large deposited energies, the crystal melts locally and then re-crystallizes. Thus, the final number of surviving defects can be much smaller than the above estimates would indicate. We are interested in very low recoil energies ($<1$\,keV); thus, this effect can be neglected.

The most accurate calculations are based on molecular dynamics (MD) simulations and fall outside the scope of our work here. Measurements and molecular dynamics calculations indicate that the idea of a sharp threshold at which the defect formation probability discontinuously jumps from 0 to 1 is not accurate. In reality, the defect creation probability evolves continuously, becoming non-zero well below the TDE and getting close to unity for values of $T_d~\sim 2-3 E_d$. Here, we employ TRIM, a binary collision simulation~\cite{TRIM}, using full cascades~\cite{Chen:2020}. Note, that the following analysis depends quantitatively on these details and the precise number of predicted vacancies as a function of energy. However, the qualitative features of the analysis are quite robust. Relative to earlier work regarding defect formation due to low-energy nuclear recoils~\cite{Budnik:2017sbu}, we would like to remark that we employ a more detailed calculation based on TRIM of defect formation in relation to recoil energies, generally resulting in higher thresholds than previously found. However, TRIM does not allow us to consider any anisotropy in response which likely will lead to a moderate increase in the variance of the number of vacancies formed at fixed recoil energy\footnote{For dark matter searches this implies that we cannot study a diurnal modulation signal.}.

The TDE has been measured for a number of elements, see, e.g., Ref.~\cite{ASTM}, but we are interested in a number of compounds for which no such data could be found. Heuristically, the TDE is found to scale with the melting point of a material~\cite{Konobeyev}. While this scaling is only approximate, it serves as a useful starting point. We find from the information given in Ref.~\cite{Konobeyev}
\begin{equation}
    E_d(T_m)=0.018\,\mathrm{eV} \left(\frac{T_m}{\mathrm{K}}\right) + 7.2\,\mathrm{eV}\,,
\end{equation}
which is obtained from a fit to 24 data points and the goodness of fit is 0.23, where $T_m$ is the melting point. This is only a rough approximation and a dedicated measurement and/or computation of this parameter is needed for the specific material in question.

The method of choice to compute the TDE is MD simulations, which also allows the study of the anisotropy of vacancy formation. Crystal damage in CaF$_2$ has been proposed for neutron dosimetry~\cite{Hecht} and, thus, been studied in detail using LAMMPS, a molecular dynamics program ~\cite{PLIMPTON19951}. In Ref.~\cite{MORRIS2020109293} the TDE averaged over all crystal directions is found to be 23\,eV, whereas, using the above heuristic, we find the TDE to be  36\,eV. This indicates that our heuristic and TRIM calculations are, overall, not very far off. Ultimately, the threshold behavior of color center formation will need to be mapped, for instance, by scattering neutrons of known energy and momentum transfer.

\subsection{General features of crystal defects from CEvNS}
\label{sec:general}

\begin{figure}
   \centering
    \includegraphics[width=\columnwidth]{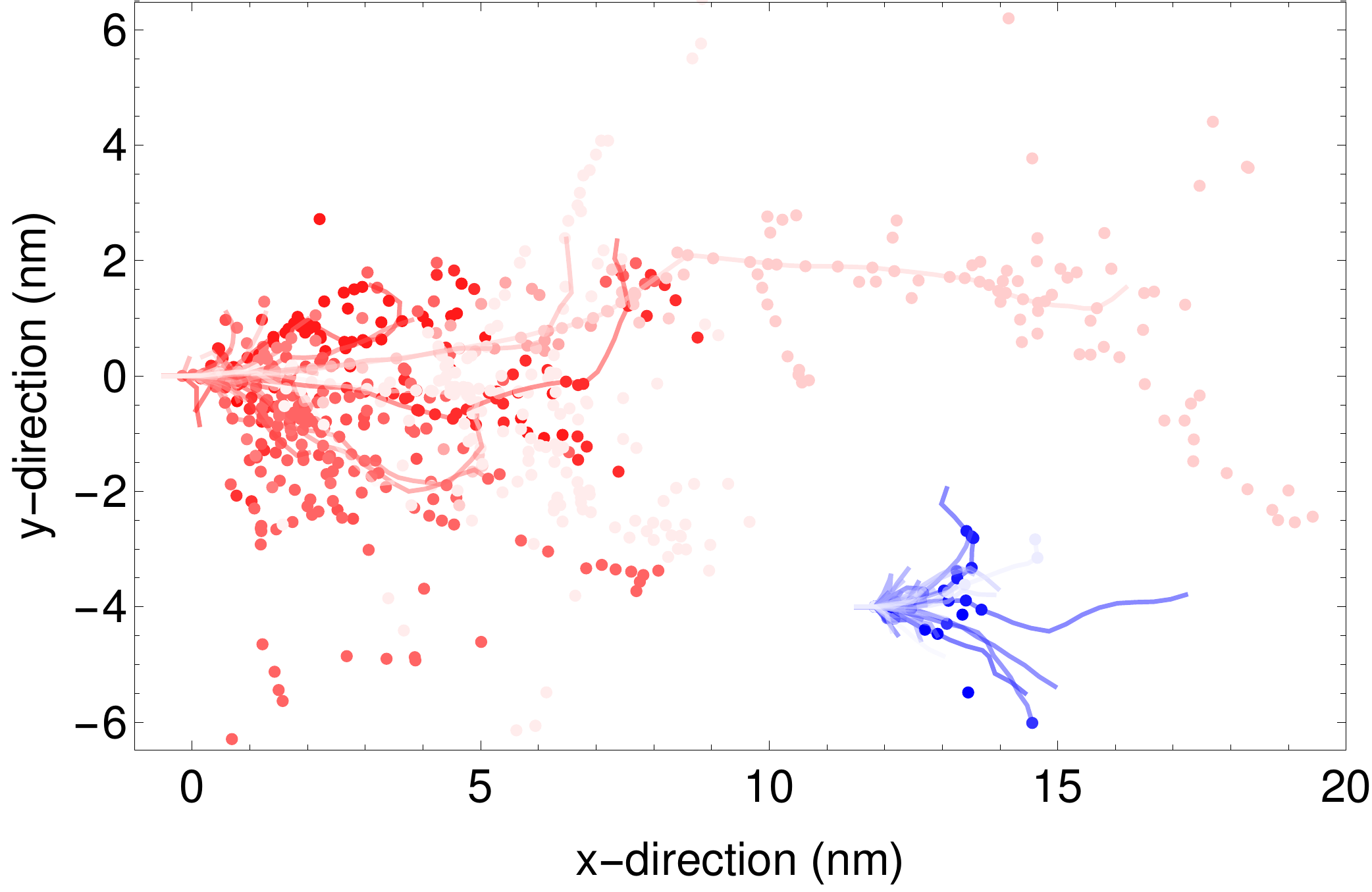}
    \caption{Shown is the overlay of 50 typical cosmic ray neutron (red) and reactor CEvNS events (blue) in NaI. Vacancies are marked by disks and tracks created by the primary recoil are marked by a line.}
    \label{fig:events}
\end{figure}

In Fig.~\ref{fig:events} we show a set of 50 neutron (red) and reactor CEvNS events (blue) in NaI, each. The lines are the tracks caused by the primary recoil and the dots are the vacancies created.  Neutron events on average are much larger spatially than CEvNS events since the mean recoil energy is large due to the $1/E$ neutron spectrum, whereas CEvNS events on average are only a few nm across since the antineutrino spectrum ends at around 8\,MeV. The rate of vacancy formation, $n_v$, is driven by the TDE, $E_d$. At large recoil energies, $E_r$ roughly scales as $n_v\propto E_r/2 E_d$. 
Coherent cross sections, either for neutrinos or dark matter, are proportional to the atomic mass squared, $A^2$, whereas, neutron scattering cross sections are essentially independent of $A$. For CEvNS, this leads to a preference for materials with a large mass fraction of high-$A$ elements, whereas, for dark matter scattering obtaining the largest possible recoil energy by minimizing $A$ is useful. Therefore, a combination of detectors, encompassing different materials with differing atomic masses, would allow the  neutron and CEvNS signatures to be statistically separated.

In terms of event rates, the sea-level primary neutron rate from cosmic rays is about one order of magnitude larger than the typical CEvNS signal close to a power reactor. The actual background rate depends on the location and details of the shielding configuration. However, the magnitude of neutron background changes relatively little as a function of the number of vacancies created, whereas, the magnitude of the CEvNS signal falls steeply with either the number of vacancies or track length. This indicates that moderate shielding combined with a focus on short tracks/few vacancies can help to reduce neutron backgrounds on a statistical basis. To obtain intrinsic rejection of gamma and beta radiation  backgrounds, we implicitly assume that ionizing radiation does not cause tracks or vacancy formation. Materials have been identified where optically active vacancies result only from nuclear recoil events~\cite{Mosbacher:2019igk}.

\begin{figure}
    \centering
    \includegraphics[width=\columnwidth]{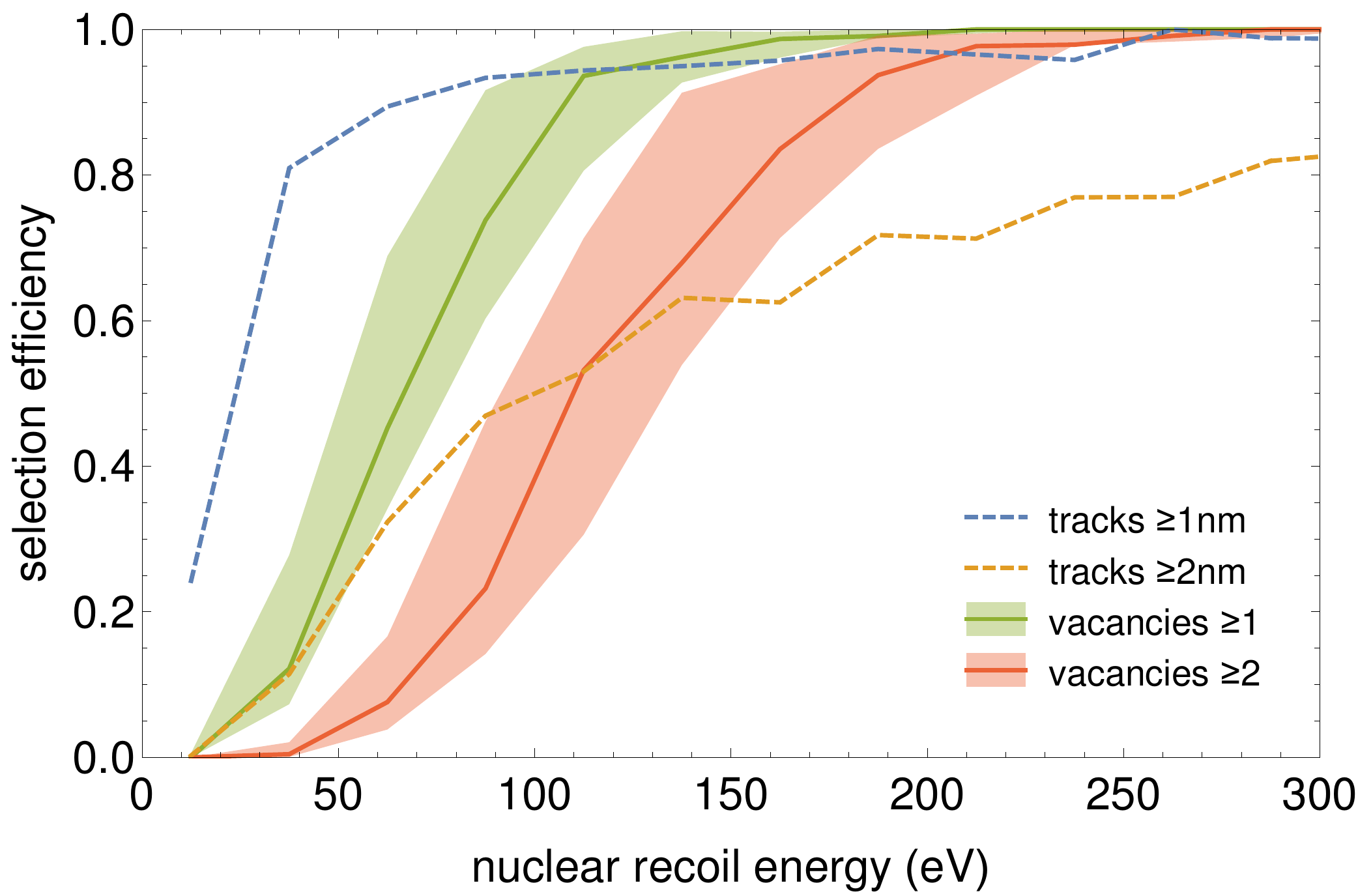}
    \caption{Shown is the nuclear recoil energy threshold in NaI for either track-based event selection (dashed lines) or vacancy-based event selection (solid lines). The bands result from a $\pm20\%$ variation of the threshold damage energy.}
    \label{fig:threshold}
\end{figure}

In Fig.~\ref{fig:threshold} we show the resulting nuclear recoil threshold behavior for track-based (dashed) event selection and vacancy-based (solid) event selection. In all cases, the nuclear recoil threshold, corresponding to 0.5 efficiency, is less than 100\,eV, which is the primary reason vacancy formation has been considered for dark matter detection~\cite{Budnik:2017sbu}. The difference between this analysis and that in Ref.~\cite{Budnik:2017sbu} is that our vacancy threshold function is based on a full TRIM calculation and we find generally higher values ($\sim 2 E_d$). Also, it becomes apparent that track-based event selection requires the ability to measure $\sim$1\,nm long tracks. The methods for track readout discussed in Refs.~\cite{Drukier_2019,Edwards_2019} are in principle sensitive to short tracks. However, the rate at which they can scan a volume decreases fast for shorter tracks. Moreover, they likely would be destroying the crystal. Therefore, overall they are not a good fit to the applications proposed here. Also, indirect track detection, via strain-induced frequency shifts in pre-existing color centers~\cite{Rajendran2017}, only applies to tracks longer than a few 10s of nm. Therefore, for our purposes, we do not further consider the track readout option.

\subsection{Readout schemes}

Color centers have been proposed as a useful detection mechanism in the context of direct dark matter searches~\cite{Budnik:2017sbu}. In particular, so-called F-centers are attractive for this application. An F-center is a vacancy of the anion in an ionic crystal which traps an electron. The resulting energy levels of this electron vacancy bound state are amenable to fluorescence, i.e., they can be excited by visible light of a suitable wavelength and the subsequent de-excitation results in the emission of photons. Individual color centers can be detected using confocal microscopy, whereby the excitation and fluorescence light share the same light path; the spatial resolution is approximately 1/2 of the wavelength~\cite{Haussler2020}. The limitation of this method is that when scanning a 3D volume only one voxel can be acquired at a given time, hence, the total volume throughput is low. In Ref.~\cite{Budnik:2017sbu} the use of an optical fiber is proposed, which effectively samples the whole volume of the fiber simultaneously, but without any ability to determine the axial position of a defect. Therefore, any pre-existing defects that are unavoidable become a background, which has to be subtracted statistically. In a dark matter context, very small (gram-scale) samples which are exposed over 100s of millions of years may still yield acceptable sensitivity. However for applications to reactors only exposure times of a few years or less are possible and hence large samples ($\sim 10-1,000$g) are required. This is where the optical, non-destructive technique we propose will become especially useful.

The work in Ref.~\cite{Budnik:2017sbu} led directly to an experimental follow-up study~\cite{Mosbacher:2019igk}: the authors measured the bulk fluorescence of a number of materials as a function of both the excitation wavelength (250--800\,nm) and fluorescence wavelength (300--1200\,nm). They compared un-irradiated samples, samples which were irradiated with gamma photons from a $^{60}$Co source with a dose  of up 1.6\,Gy and samples which were exposed to neutrons from a $^{252}$Cf source with a fluence of roughly $10^{10}\,\mathrm{cm}^{-2}$ corresponding to a kerma of about 0.1\,Gy. For some materials it could be shown that color centers are selectively formed only by neutron irradiation and not by gamma irradiation. These materials and their excitation/fluorescence wavelengths are: Al$_2$O$_3$ (250,400)\,nm; SiO$_4$ (250,346)\,nm; LiF (260,331)\,nm; BaF$_2$ (250,270)\,nm. Note, that both LiF and BaF$_2$ are candidate materials for the technique discussed here, as explained in detail in Sec.~\ref{sec:materials}. Moreover, this data allows to put an upper limit on the intrinsic, pre-exisiting color center density in those materials at the $10^{-12}-10^{-10}$ level (see the Appendix). No other optical backgrounds are apparent from this data. We also learn from this data, that color centers with different excitation/fluorescence wavelengths combinations exist within the same material and some do respond to gamma radiation and some do not.

\begin{figure*}
    \centering
    \includegraphics[width=\textwidth]{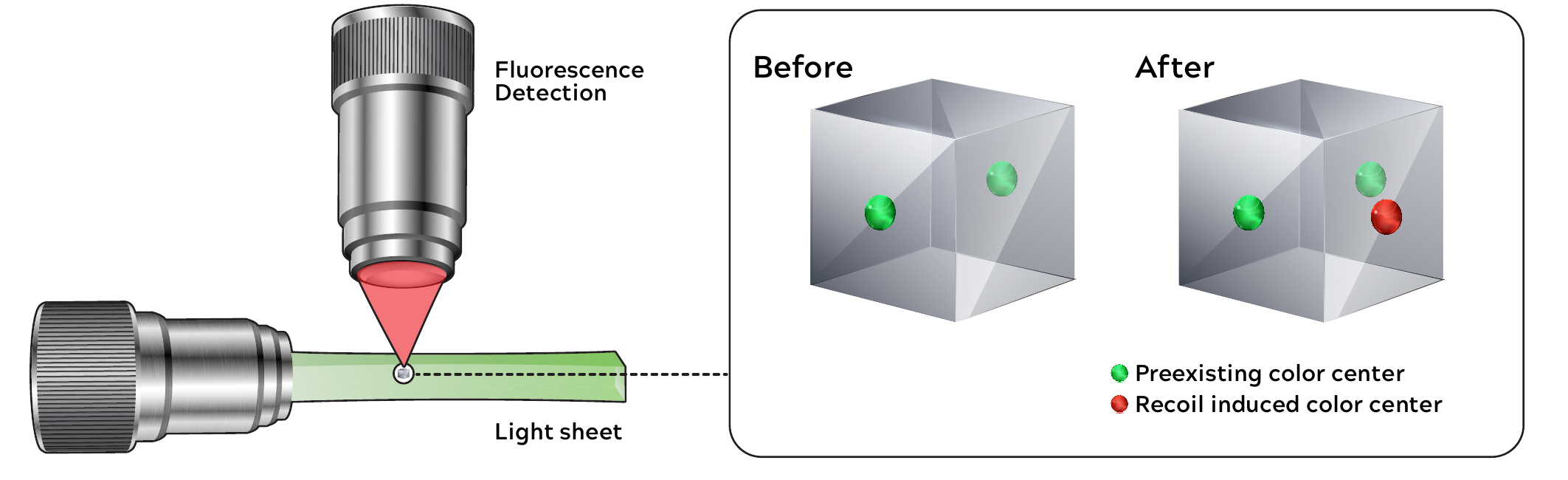}
    \caption{Schematic layout of light-sheet microscopy used for imaging single color centers pre- and post-exposure.}
    \label{fig:schematic}
\end{figure*}

We propose the use of light-sheet microscopy, also known as Selective Plane Illumination Microscopy (SPIM)~\cite{Huisken1007}, which is an evolution of confocal microscopy, wherein the excitation light and fluorescence light paths are orthogonal to each other. A sheet of light illuminates the entire field of view of the fluorescence light path and, thus, as many voxels as there are pixels in the camera along the fluorescence light arm can be imaged simultaneously, see Fig.~\ref{fig:schematic}. For a review of this technique, see, e.g., Ref.~\cite{lightsheet}. For biological systems there is an interest in observing dynamic processes in whole organisms; hence, repeated fast sampling of a relevant volume is desired. One potential technology for this purpose is spherical-aberration-assisted extended depth-of-field (SPED) light-sheet microscopy~\cite{Tomer2015}. The point spread function of the fluorescence light path is extended so that a large volume can be imaged. With SPED, imaging of 12 volumes of $0.9\times0.4\times0.2\,\mathrm{mm}^3$ per second has been demonstrated, which translates to 75$\,\mathrm{cm}^3$ per day. This volume is scanned in 40 steps in the z-direction, making each z-slice $5\,\mu$m thick and the time per slice is $1/(40\times12)\simeq 2\,\mathrm{ms}$. Hence, this technique allows volume sampling in the range of 50--100$\,\mathrm{cm}^3$ per day. 

Assuming $\mathcal{O}(1)$ pre-exisiting defects per voxel (corresponding to a defect density of $10^{-10}$ defects per atom, which is the upper limit derived from the data in Ref.~\cite{Mosbacher:2019igk}), only a relatively small number of fluorescence photons need to be acquired to obtain positive detection of a new defect. To achieve a false positive defect detection rate of $10^{-12}$, that is, roughly only one false positive for the entire detector, and assuming 1 pre-existing color center per voxel, about 100 photons are required to conclusively detect the presence of one additional defect. Supposing an oscillator strength of order 1 for the color center, we end up with an absorption cross section of $\sim 10^{-17}\,\mathrm{cm}^2$~\cite{Budnik:2017sbu}. Accounting for quantum efficiency of the camera and the light collection efficiency of the objective we find the following parameters for scanning 100\,g  of BaF$_2$: a laser power of 12\,W, a temperature rise in the 1\,mm thick sample of 3.4\,K and a data volume of 41\,TByte; for details see the Appendix, where we also give the appropriate scaling laws. From this discussion it follows that the number of pre-existing color centers of the type induced by nuclear recoil is a key figure of merit for any material and will need to be experimentally determined.  

Moreover, the crystal could be imaged prior to deployment and, thus, a voxel-by-voxel subtraction of background is feasible, as shown in  Fig.~\ref{fig:schematic}  At the same time the comparison of  pre- and post-deployment images serves to verify the identity of the crystal and that it has not been tampered with, by for instance heating it to anneal defects. These features, in combination, allow for single event detection.

\subsection{Materials Selection}

\label{sec:materials}

Existing studies and physics considerations help identify 6 material candidates. In terms of material properties, we propose that acceptable candidates meet the following criteria:
\begin{itemize}
    \item Have a melting point well above room temperature
    \item Be an electrical insulator
    \item Permit optically active defects (color centers)
    \item Have color centers which are selectively formed by nuclear recoils, only.
    \item Be able to produce optical quality crystals
    \item Contain high-mass (for CEvNS) or low-mass (for dark matter and neutrons) elements
    \item Have a low threshold damage energy (TDE)
    
\end{itemize}

\begin{table*}[]
    \centering
    \begin{tabular}{c|rrrrrrrrrr}
         & $A$ & $m_A$ & density & melting& TDE & $E_\mathrm{1/2}$&CEvNS& selection& usable  \\
        &&&&point&&&events&efficiency&events\\
        material& [u]&[\%]&[$\mathrm{g}\,\mathrm{cm}^{-3}$] & [K] &[eV] &[eV]&&[\%]&\\
        \hline
        \hline
         LiF & 19.0 & 73.2& 2.64 & 1120 & 27 & 80&5600&75&4200\\
         BaF$_2$& 137.3 & 78.3 & 4.88 & 1625 & 35 & 105 & 48800 & 20 & 9600\\
         NaI & 126.9 & 85.0 & 3.67 & 935 & 24&65&46900&32&15100\\
         CsI & 132.9 & $\sim$100 & 4.51 & 900 & 23 & 55 & 55700 & 37 & 20600\\
         CaWO$_4$ & 183.8 & 63.9 & 6.06 & 1895 & 41 & 110 & 55400 & 8 &4600\\
         Bi$_{12}$GeO$_{20}$ & 209.0 & 86.4 &9.22 & 1175 & 28& 85 &83500 & 17 & 14000\\
    \end{tabular}
    \caption{Materials considered in this study. $A$ is the atomic mass of the heaviest element and $m_A$ the corresponding mass fraction. TDE is threshold damage energy, $E_\mathrm{1/2}$ is the nuclear recoil energy at which 1/2 of the asymptotic efficiency is reached. CEvNS event numbers are computed for 100\,g of target and 1 year exposure at 10\,m from a 3\,GW$_\mathrm{th}$ reactor. The same number for inverse beta decay on CH$_2$ is 1250 . Selection efficiency is the fraction of reactor CEvNS events which have one or more vacancies created. Usable events is the number of events where one or more vacancies are created.}
    \label{tab:materials}
\end{table*}

Color centers, i.e., crystal defects that are optically active, are common and were first discovered in rock salt (NaCl). They are known to exist in all chemical group I-VIII compounds. Color centers are not confined to this group of compounds, however, and are found in many materials, including glasses. For LiF and BaF$_2$ color centers have been confirmed to be selectively created only by neutron and not gamma irradiation~\cite{Mosbacher:2019igk}. That is, LiF and BaF$_2$ detectors will be immune to ionizing radiation. The cited study of these two materials was specifically performed with the goal of dark matter detection. Also, color centers in NaI~\cite{nai}, CsI~\cite{csi}, CaWO$_4$~\cite{cawo4}, BGO~\cite{bgo} have also been observed in response to X-ray and electron irradiation. However, specific studies to determine if a neutron/gamma separation, like in LiF, is possible are required for these other 4 proposed materials. 

The neutron cross section is more or less independent of the atomic mass, $A$, whereas, the CEvNS cross section scales with $A^2$. Therefore, heavy elements are preferred. Within the group I-VIII compounds, this makes NaI and CsI prime candidates. Therefore, in terms of finding an ideal material, we look for the heaviest possible elements and materials with the largest possible concentration of those heavy elements. Two candidates that fit this description are $\mathrm{Ca}\mathrm{W}\mathrm{O}_4$ and $\mathrm{Bi}_{12}\mathrm{Ge}\mathrm{O}_{20}$ (BGO).  Both materials appear in nuclear and particle physics applications as scintillators and/or low-temperature bolometers. This implies their ready availability (and relative affordability) as large (100s of grams) crystals with good optical qualities and a low radioactive impurity content.  With bismuth and tungsten being at the upper range of atomic masses for stable elements, and given the large weight fraction of these in the above compounds, it at first appears, theoretically, that these two materials would be the most attractive target materials.

The results from the TRIM simulation, for the 6 materials identified, are shown in Tab.~\ref{tab:materials}. While the simple argument given above bears out for the actual number of CEvNS events, looking at the fraction of reactor CEvNS events that produce one or more vacancies, we can see that the number of actual usable events does not follow this simple trend. LiF does remarkably well in this metric compared to CaWO$_4$. LiF has a fairly high value of $E_{1/2}$ but also the mean recoil energy from reactor CEvNS events is large, so that the selection efficiency is 75\%. For CaWO$_4$ on the other hand, the mean  recoil energy from reactor CEvNS events
is small and, hence, the selection efficiency is only 8\%. The selection efficiency for one or more vacancies created is well below unity, which would be much less of a problem in a track-based analysis, i.e., many more  events leave tracks of 1\,nm or longer. However, the neutron background rises very steeply at the lowest energies and, thus, these very low energy events are not very useful. We confirmed this by computing the same results as shown in Figs.~\ref{fig:sigma} and~\ref{fig:safeguards} with a track based selection for events (with tracks of 1\,nm or longer) and obtained virtually identical results. Given our inability to find a convincing readout scheme for tracks that short, we do not include these results here. Note also, that high-mass materials offer a better distinction of their CEvNS events from neutrons since their signal peaks at low energies, that is few events will create multiple vacancies, whereas, the neutron vacancy distribution is nearly independent of the atomic mass.  For the same reason, LiF, despite its usable event number being comparable to CaWO$_4$, does quite poorly as a reactor CEvNS detector; the shapes of the signal and background distributions are too similar.

We would also like to remark that the results of Ref.~\cite{Bowen:2020unj} are confirmed in our analysis: thresholds of a few 100\,eV or less are needed to gain any advantage relative to inverse beta decay in event rate per unit mass. For the six materials in Tab.~\ref{tab:materials}, we find a mass advantage on usable events of 3.3--16. Mass advantage is defined as the ratio of the mass of an IBD-based detector to the mass of a CEvNS-based detector, at the same performance~\cite{Bowen:2020unj}.  

For dark matter detection the considerations are somewhat different. Dark matter detectors are located deep underground. Consequently, neutron backgrounds are of less concern. Neutron background-free measurements up to an exposure of 10s of kg-years have been reported (see e.g., Ref.~\cite{Behnke2017}). As we will discuss in detail later, LiF is an ideal material because of its very low atomic mass and because both Li and F have a net nuclear spin.

\section{Results}
\label{sec:results}

\subsection{Dark matter detection}

There is ample evidence for dark matter based on the dynamics of galaxies and galaxy clusters as well as from precision cosmology~\cite{Battaglieri:2017aum}. Assuming that some new species of particles is responsible for these results, a picture emerges where the Milky Way, like all other galaxies, is embedded in a dark matter halo and the Sun's orbit around the center of the Milky Way leads to an effective dark matter wind. Dark matter being dark, i.e., electrically neutral, the most plausible detection mechanism is the creation of nuclear recoils via some new interaction~\cite{Goodman:1984dc}. We broadly follow the description in Refs.~\cite{Lewin:1995rx,Schnee:2011ooa} in order to compute the expected event rate for dark matter scattering in our detectors as well as the recoil energy spectrum. The detection mode discussed here is relevant for spin-dependent interactions and given as~\cite{Schnee:2011ooa}
\begin{equation}
    \frac{d\sigma_\chi}{dq^2}=\frac{F^2(q)}{4\mu_A^2 v^2}
    \frac{32G_F^2\mu_A^2}{\pi}\frac{J+1}{J}\left(a_p\langle S_p\rangle+a_n\langle S_n\rangle\right)^2\,,
\end{equation}
where $\mu_A$ is the reduced mass of the colliding system, $a_{p/n}$ are the spin-dependent coupling to the proton and neutron, respectively, $J$ is the total nuclear spin, and $\langle S_{p/n}\rangle$ are the average proton and neutron spins in the nucleus. $F^2(q)$ is the nuclear form factor, which we set to 1 since we are interested in sub-keV recoils only. To evaluate the spin-dependent cross section we use for $^{19}$F the proton-coupling $\langle S_p\rangle=0.477$~\cite{Schnee:2011ooa}, for $^7$Li we use  $\langle S_p\rangle =0.497$~\cite{Abdelhameed:2019szb} and we set  $\langle {S_n} \rangle\simeq0$ for both isotopes.  Here, $v$ is the velocity of the dark matter particle in the laboratory frame. Therefore, this cross section needs to be integrated with respect to the dark matter velocity distribution, which is assumed to follow a truncated Maxwell distribution with  $v_0=230\,\mathrm{km}\,\mathrm{s}^{-1}$ and a cut-off of $v_\mathrm{esc}=600\,\mathrm{km}\,\mathrm{s}^{-1}$.
Also, we use $\rho_0=0.4\,\mathrm{GeV}\,\mathrm{cm}^{-3}$.  The maximum nuclear recoil energy is given by
\begin{equation}
    E_\mathrm{max}=\frac{4 M_\chi M_A}{(M_\chi+M_A)^2}\frac{1}{2}M_\chi v_\mathrm{esc}^2\,,
\end{equation}
which yields $E_\mathrm{max}\simeq40$\,eV for $A=7$ and $M_\chi=0.25\,\mathrm{GeV}$. For spin-independent dark matter couplings, our detection scheme is unlikely to provide a relevant advantage since very low-threshold silicon (DAMIC, SENSEI) and germanium detectors (SuperCDMS) are available already~\cite{Aguilar-Arevalo:2019jlr,SENSEI:2020dpa,SuperCDMS:2020ymb}

\begin{figure}
    \centering
    \includegraphics[width=\columnwidth]{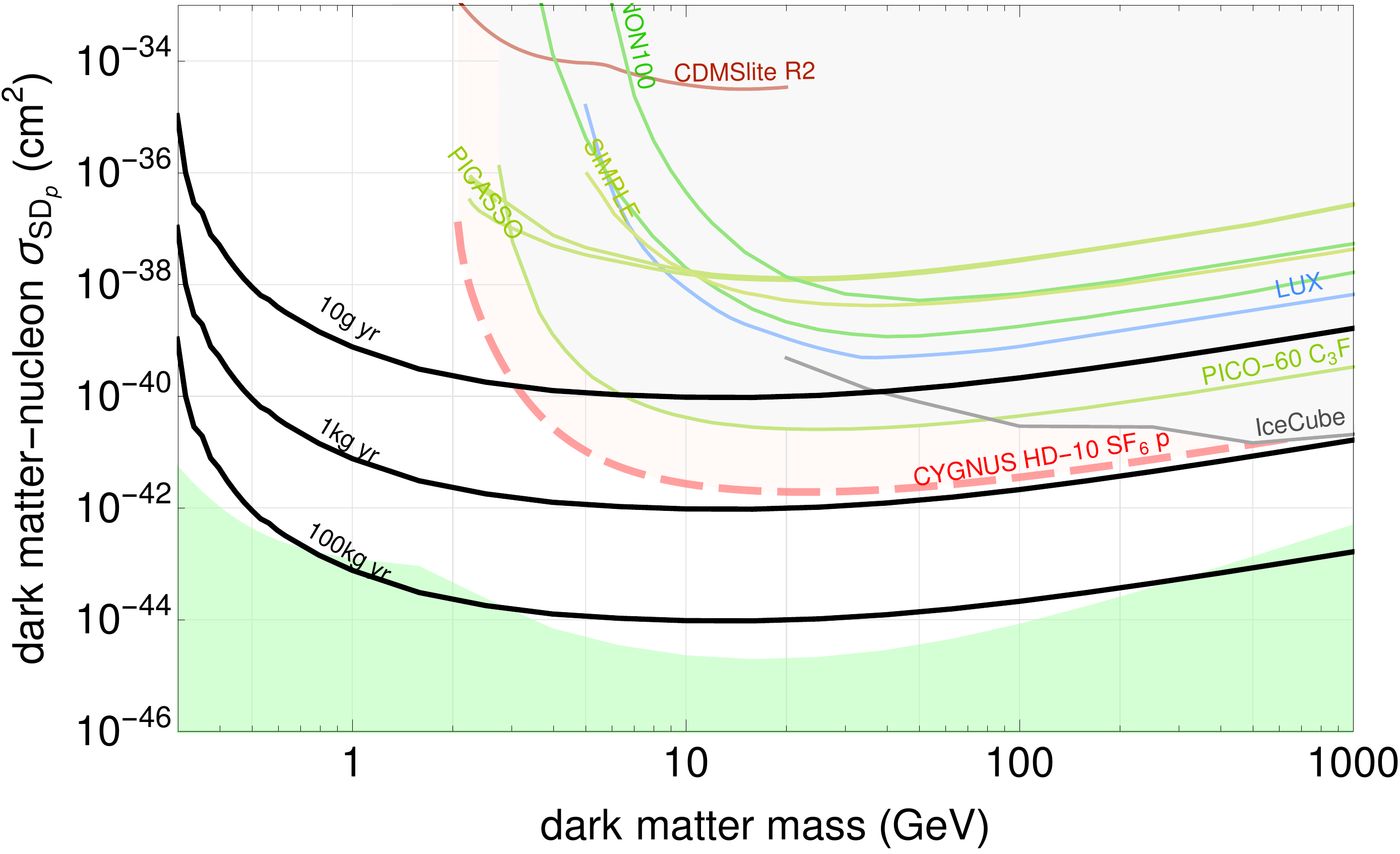}
    \caption{Shown is the 95\% CL sensitivity with LiF for spin-dependent dark matter scattering (black lines) for different exposures assuming a background free experiment. The green shaded region is the neutrino floor for LiF. All other experiments and exclusion limits have been plotted using~\cite{dmplot}, specifically SIMPLE II~\cite{Felizardo2012}, PICO-60~\cite{Behnke2017}, LUX~\cite{Akerib_2017}, Panda-X II~\cite{Fu2017}, XENON-100~\cite{Aprile2016}, CDMSlite~\cite{cdmslite}. }
    \label{fig:darkmatter}
\end{figure}

To optimize sensitivity to spin-dependent dark matter couplings, nuclei with a net total nuclear spin are required. To optimize sensitivity to small dark matter masses, light nuclei are preferred. This makes LiF a good candidate material~\cite{Mosbacher:2019igk}. Both $^7$Li and $^{19}$F have a net nuclear spin of 1/2 and sizable average proton spins. They are also both rather light. $^{19}$F is the single stable isotope of fluorine and $^7$Li has an isotopic abundance of 92.4\%. Consequently, both elements have been used to set limits on spin-dependent dark matter couplings~\cite{Abdelhameed:2019szb,Amole_2019}. In Fig.~\ref{fig:darkmatter} we show the results for various exposures of LiF, where our exclusion limit is defined by requiring 2.3 events. We assume that the experiment is otherwise background free. Given the experience of the PICO-60 experiment ~\cite{Behnke2017,Abdelhameed:2019szb} and extrapolation to PICO-500~\cite{olson}, our assumption seems plausible for target masses up to 10 kilograms even in the absence of a muon veto.

In producing these limits we compute the recoil spectrum for a given dark matter mass for both Li and F and use TRIM to compute the resulting  number of vacancies. Hence, we  can select all events which have one or more vacancies.  The neutrino floor is computed using neutrino fluxes for solar and atmospheric neutrinos~\cite{Vitagliano:2019yzm} as described in Ref.~\cite{Billard:2013qya}. This neutrino floor would be reached for exposures of around 100\,kg\,-yr. Given an estimated readout rate of 100-200\,g per day, exposures above $\sim10$\,kg\,-yr will require scaling up the readout rate by a factor of a few, which is probably best accomplished by using several microscopes in parallel. Note, that the readout for the dark matter detection has to happen underground, because neutron backgrounds at the surface would overwhelm any signal in this case. The resulting exclusion limits we find in our analysis are unmatched for dark matter masses below 3\,GeV and extend down to 0.3\,GeV representing a one order of magnitude improvement in mass reach.

\subsection{Reactor CEvNS}

Coherent elastic neutrino-nucleus scattering (CEvNS)  occurs when a neutrino interacts coherently with the total weak nuclear charge (necessarily at low momentum transfer) leaving the ground state nucleus to recoil elastically~\cite{Freedman:1973yd}. It was observed for the first time in 2017 with 50\,MeV neutrinos from the Spallation Neutron Source (SNS) at Oak Ridge~\cite{Akimov:2017ade}.

The CEvNS cross section is enhanced by $N^2$, with $N$ being the neutron number and, thus, can exceed inverse beta decay (IBD) cross sections by more than 2 orders of magnitude per unit detector mass. However, this advantage can only be realized for very small recoil energy detection thresholds~\cite{Bowen:2020unj}. The SM cross section for coherent elastic neutrino scattering is given by~\cite{Freedman:1973yd}

\begin{equation}
\label{eq:crosssection}
        \frac{d\sigma}{dT}(E_\nu,T) = \frac{G_F^2}{4\pi}N^2 M \left(1 - \frac{M T}{2E_\nu^2} \right)\,,
\end{equation}
where  $G_F$ is the Fermi constant, $M$ is the total mass of the nucleus, $T$ is the nuclear recoil energy and $E_\nu$ is the neutrino energy. Due to the low energy of reactor neutrinos, we can safely set the nuclear form factor to 1. For a given neutrino energy, kinematics limits the recoil energy to be less than
\begin{equation}\label{eq:tmax}
    T_\mathrm{max}=\frac{E_\nu}{1+\frac{M}{2E_\nu}}
\end{equation}
which for a target mass $A=100$ and $E_\nu=8$\,MeV equates to about 1200\,eV and for $E_\nu=1\,$MeV to about 20\,eV, which are typical energy endpoints for reactor neutrino fluxes. These very small nuclear recoil energies are the reason that CEvNS has not yet been observed at reactors. The number of interactions is given by
\begin{equation}\label{eq:eventnumber}
    n=\frac{G_F^2}{4\pi}M\int_T^{Tmax}\int_0^{\infty}dE_{\nu} \cdot \phi(E_{\nu})  N^2 \left(1 - \frac{M T}{2E_\nu^2} \right)\,,
\end{equation}
where $\phi(E_\nu)$ is the reactor neutrino flux, which we obtain from a summation calculation performed following the formalism outlined in Ref.~\cite{Huber:2011wv} down to a neutrino energy of 250\,keV. For simplicity, we assume a reactor running on pure $^{235}$U. More realistic fission fractions would change event numbers by less than 20\%. Neutrinos from neutron capture reactions on $^{238}$U~\cite{Cogswell:2016aog} cause recoils that are below the energy threshold considered here and can be safely neglected.

There are a large number of experiments under way aiming to be the first to detect CEvNS at a reactor using more complex and higher-maintenance methods than the one proposed here: CONUS using cryogenic (70\,K) germanium detectors~\cite{Bonet:2020awv}, MINER  using cryogenic (15\,mK) germanium detectors~\cite{Agnolet:2016zir}, 
NUCLEUS based on cryogenic (15\,mK) CaWO$_4$ bolometers~\cite{Rothe:2019aii}, RED-100 a dual-phase liquid xenon (170\,K) detector~\cite{Akimov:2019ogx},  NEON NaI scintillator detectors~\cite{Choi:2020gkm}, RICOCHET low-temperature bolometers (15\,mK) ~\cite{Billard:2016giu}, TEXONO cryogenic germanium (70\,K) detectors~\cite{Wong:2016lmb}, CONNIE using silicon CCDs at 100\,K~\cite{Aguilar-Arevalo:2019jlr}. Notably, only NEON operates at room temperature. However, the only experiments to demonstrate nuclear recoil thresholds below 100\,eV are bolometers operating at mK temperatures. 

In light of the current lack of understanding of recoil background at the low energies relevant here, we would like to emphasize, that the detection method we propose is in general not more or less affected by a genuine neutron backgrounds than existing technologies. The one exception being muon-induced events, which in principle can be reduced using an active muon veto. Since this requires event timing information, which our technology can not provide, this option is not available. On the other hand, should the observed low-energy excess~\cite{excess} be due to yet to be understood energy depositions in the detector not related to nuclear recoil, the technology we propose here could help to disentangle these unknown contributions. Existing detector technologies ultimately measure an energy deposition, which in principle can have many sources, in our proposal we measure the creation of very specific crystal defects in the form of color centers at a microscopic level.

Cosmic ray neutron backgrounds cannot be avoided for a close-to surface deployment. Neutron recoils produce exactly the same type of recoil as the CEvNS reaction and, thus, are the leading source of background. For our purposes, recoils with energies in the 100--1,000\,eV range are most relevant. Hence, neutrons in the 100--10,000\,eV range cause the relevant background events. Neutrons at the surface of the earth are predominantly  produced by cosmic ray interactions within the atmosphere.  Cosmic ray muons interacting in materials close to the detector can also produce neutrons. In practice a fine balance between these two background sources will need to be achieved.  For shielding exceeding a few meters of water equivalent muon interactions in the shielding quickly become the dominating source of neutrons. A detailed understanding of neutron backgrounds and the effectiveness of shielding will require a careful simulation of the specific setup and materials. Ultimately, only an actual measurement can settle the issue.

Neutrons in the energy range  of concern, 100--10,000\,eV, can be effectively shielded in surface environments. Shielding factors of up to 100 have been demonstrated with shields of an areal density less than 100\,$\mathrm{g}\,\mathrm{cm}^{-2}$~\cite{bergevin2019applied}. A similar result is found in the context of upcoming and ongoing CEvNS experiments at reactors~\cite{Bonet:2020awv,conus,nucleus,connie, richochet,MINER}: typical background rates at the surface of around 10,000 events per day and kilogram of detector in the energy range from  20--10,000\,eV are stipulated. This agrees well with our simplified model. Therefore, it appears at least possible that passive shielding could yield a suppression factor of 100.  Note that, active shielding would require the detector to be read out in real time and thus is not considered here. On the other hand, should the low-energy event excesses reported by a number of experiment indeed be due to genuine recoil events, it would be unclear if backgrounds can be overcome by this, or any other, technique.

\begin{figure}
    \centering
    \includegraphics[width=\columnwidth]{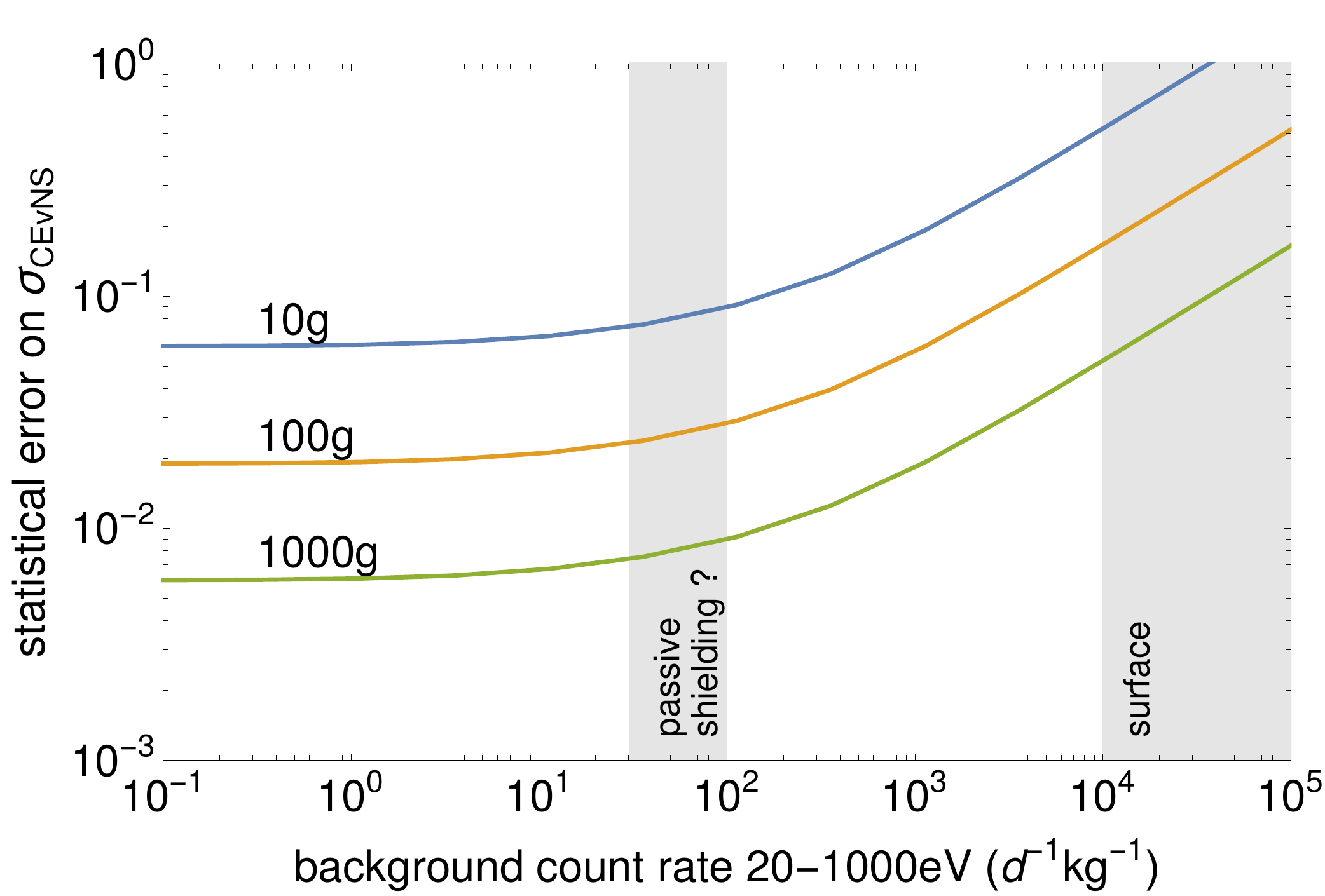}
    \caption{Shown is the 1\,$\sigma$ error on the CEvNS cross section obtainable at a nuclear reactor for a variety of detector masses in a vacancy-based counting analysis, as a function of the neutron background rate. The exposure is one year at 20\,m from a 3\,GW$_\mathrm{th}$ reactor.}
    \label{fig:sigma}
\end{figure}

In Fig.~\ref{fig:sigma} we show the 1\,$\sigma$ error on a measurement of the CEvNS rate at 20\,m from a 3\,GW$_\mathrm{th}$ reactor for a 1 year deployment as a function of the neutron background rate. The assumption is made that no reactor related background from fast neutrons is present. The data and background recoil rate and spectra are computed as described above and the ensuing recoil spectra are put into TRIM, which then outputs the distribution of produced color centers. The resulting signal and background histograms are used to compute the 1\,$\sigma$ error using a maximum likelihood method. The bin with zero color centers formed is excluded in the analysis, resulting in an overall 32\% efficiency (see Tab.~\ref{tab:materials}). In evaluating the likelihood, the normalization of the background is treated as an unconstrained nuisance parameter and profiled over. The result is statistical only, that is, no analysis on reactor flux systematics or signal normalization is included.

\begin{figure}
    \centering
    \includegraphics[width=\columnwidth]{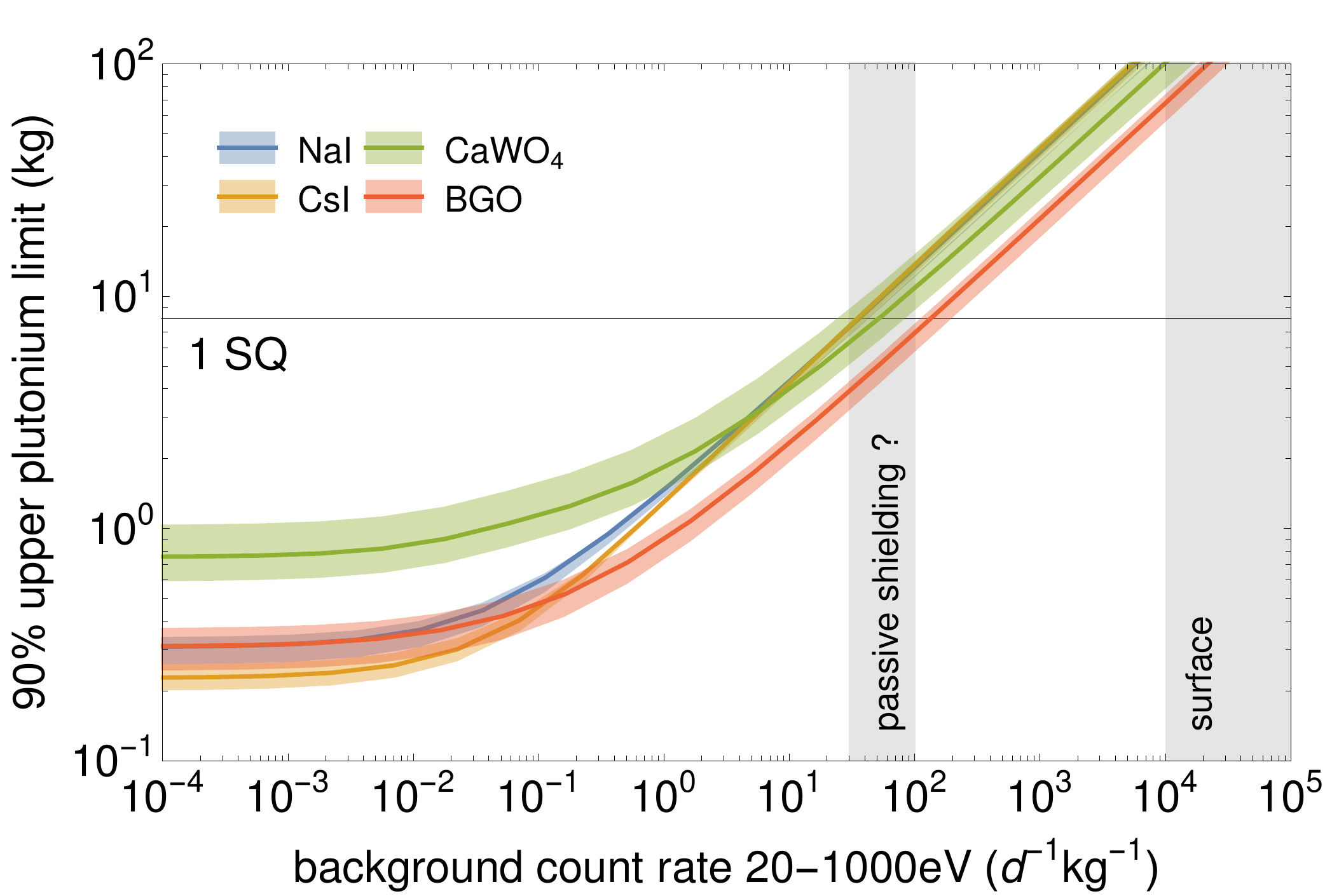}
    \caption{Shown is the 90\% CL limit upper bound on the amount of produced plutonium for a  reactor which is shut down as a function of the neutron background rate. The detector size is 100\,g, the distance is 20\,m and the data taking period is 90 days. The different curves are for different materials, as specified in the legend, and the width of the band is due to a $\pm20\%$ variation of threshold damage energy. The black horizontal line indicates  8\,kg of plutonium, the so-called significant quantity (SQ).}
    \label{fig:safeguards}
\end{figure}

Even a 10\,g crystal with moderate shielding could easily achieve a 5\,$\sigma$ detection and 100--1,000\,g detectors can probe the physics associated with CEvNS well into the regime where reactor flux systematics will become the limiting factor. NaI is used an example, but many other materials would lead to comparable results (see also Fig.~\ref{fig:safeguards}). To distinguish the CEvNS signal from backgrounds three handles are available: the vacancy number distribution, with neutrons producing a higher mean number of vacancies; the reactor on/off comparison; and co-deploying a high atomic mass crystal like BGO with a low atomic mass crystal like LiF would exploit the different scaling with atomic mass of signal and background to uniquely establish the signal as stemming from CEvNS. The materials considered in our study span the atomic mass range from $A=7-209$, which is also invaluable for testing and discriminating various new physics models.

Neutrino applications to reactor monitoring and nuclear security were suggested by Borovoi and Mikaelyan ~\cite{Borovoi:1978} around the same time CEvNS was proposed. Since then, there have been many studies conducted and detectors built with the goal of real-world implementation~\cite{Bernstein:2019hix}. However, roadblocks arise from the complexity of neutrino detectors, which are still more like a science experiment than a field-deployable robust device, and the need to deal with cosmic ray backgrounds by going underground. Major excavation is generally considered impossible close to a reactor and only a few reactor sites have an existing suitable underground site. In 2018, the problem of underground neutrino detector deployment was solved by the first demonstration of surface reactor neutrino detection~\cite{Ashenfelter:2018iov,Haghighat:2018mve}.

The Treaty on the Non-Proliferation of Nuclear Weapons (NPT) is the backbone of efforts in nuclear security and nonproliferation. It includes an extensive verification regime, which is implemented by the International Atomic Energy Agency (IAEA). The IAEA is faced with the task of inspecting more than 400 commercial reactors, and around double that number of research reactors, on a limited and relatively fixed budget. Therefore, a typical neutrino detector, with its cost, ton-scale and several square meters of footprint, floor loading, maintenance requirements, etc., is not a particularly good fit for the IAEA~\cite{IAEA2012}.

The detector type we propose here can, in principle, overcome many of these IAEA implementation constraints. They are small (sub-kilogram), cheap,  robust, passive, operate at room-temperature, maintenance-free and do not present a fire hazard. Overall  the concept of operations is very similar to tags and seals, which the IAEA already uses, wherein an IAEA inspector would emplace the detector and use a seal to ensure it stays secure. After a period of time, they would return, retrieve the detector and ship it to a central laboratory for analysis ~\cite{IAEA2011}. A scan of pre-existing defects, prior to deployment, would allow for unique identification of the detector crystal when it returns for analysis after a deployment cycle.

This pre-and post-imaging analysis can also verify that the detector has not been tampered with. For instance, heating it in an attempt to anneal crystal defects would also reduce the number of pre-existing defects and thus be detectable. One scanning station could be used for a large number of detectors.  Also, being entirely passive, 
these detectors are immune from cyber-threats. In the case of a cyber-attack on conventional systems, they could provide continuity of knowledge or a redundant backup in case of a power outage induced by malicious or climate-change based catastrophic events. Seals and other passive techniques currently used by the IAEA share this trait, but passive CEvNS detectors can additionally provide direct information on reactor operations.  Each detector would essentially be only as costly as its materials. Since the materials discussed here are all commercially available, this cost would be relatively low.

In Fig.~\ref{fig:safeguards} we show the limit on plutonium production in a reactor that can be obtained with 100\,g of various detector materials as a function of the neutron background rate, at a distance of 20\,m from reactor output, and within a data collection time of 90 days. The detection goal of the IAEA for plutonium is 1 significant quantity (SQ), defined as 8\,kg, within 90 days at 90\% probability~\cite{IAEA2001}. The horizontal line labeled "1 SQ" indicates this limit. The width of the bands arises from a $\pm20\%$ variation of TDE as given in Tab.~\ref{tab:materials}. There is a range of materials which, with moderate shielding, can meet this criterion. Since the detector itself is small, an appropriate shield of about 0.5-1\,m thickness still yields a compact device. And the passive shield does not need to be moved when the detector is retrieved; it only needs to be put permanently in place once.

Outside of the reactor building (to which a 20\,m distance roughly corresponds), a small diameter, $\sim 10\,$cm, shallow borehole of 1-2 meter depth can easily provide the requisite shielding for this passive CEvNS detection method; as could a simple barrel full of water. Co-deploying a low-atomic mass detection medium like LiF can serve as an independent background assay. Compared to neutrino detectors based on inverse beta decay, the only ones demonstrated to date, the information content of the signal of the detector proposed here is a rate-only measurement with no time information. That is, our detectors provide measurements of the total neutrino fluence over the deployment period, with no information on the neutrino energy spectrum. Thus, use cases related to determining the fission fractions in a reactor (see, e.g., Ref.~\cite{Christensen:2014pva}), are not covered.

On the other hand, use cases that only need a total rate measurement can be effectively addressed with the proposed passive CEvNS method; in particular, since the event rate per unit mass is favorable for CEvNS given the very low recoil threshold we can achieve~\cite{Bowen:2020unj}. One potential use case would be the verification of reactor shutdown and, hence, absence of plutonium production during a given period. Another use case could be the monitoring of larger research reactors to ensure that the product of reactor power and operation time stays below a certain bound, to exclude their use for plutonium production. For this application the IAEA is currently using the so-called thermohydraulic power monitor~\cite{atpm}, which is used to measure the flow rate of coolant and temperature rise across the reactor. In comparison, the detectors proposed here would be cheaper, easier to install and require less maintenance.

\section{Summary \& Outlook}
\label{sec:summary}

In this paper we propose using crystal damage caused by nuclear recoils for dark matter detection and CEvNS detection with reactor neutrinos. We call this technique PALEOCCeNe. We perform a detailed simulation of track formation and vacancy creation, using the TRIM software, and present results for six promising candidate materials. In the selected materials, vacancy creation eventually results in the formation of color centers. We propose a readout scheme using  selective plane illumination microscopy, which exploits the fluorescence of these color centers. This specific type of microscopy can image 50-100\,cm$^3$ per day at sub-micron resolution with single color center sensitivity and has been developed for applications in biology. This also enables  imaging of pre-existing (unavoidable) color centers. This data can then be used for event-by-event discrimination from color centers created during detector deployment.

This is a major improvement relative to the color-center-based dark matter and solar neutrino detection scheme put forward previously in Ref.~\cite{Budnik:2017sbu}. 
In a nuclear security application the pattern of pre-existing color centers also serves as a unique identification of the specific detector deployed~\cite{Philippe2020} and as a means to identify detector tampering. We also discuss the cosmic ray neutron background for reactor measurements. Neutron backgrounds can be determined in situ by either exploiting color center multiplicity or by co-deployment of a low-atomic mass detector and leveraging the $A^2$ dependence of CEvNS.
We use this analysis framework to study three specific applications: 

\begin{enumerate}
\item Direct dark matter detection with spin-dependent forces in LiF: A 10\,g-yr exposure would set the world's best limit for dark matter masses between 0.3--3\,GeV. Scaled to a 1\,kg-yr exposure this detector would set the best limit up to masses of 1,000\,GeV. 
\item Reactor neutrino CEvNS discovery in NaI: A 10\,g-yr exposure at 20\,m from a 3\,GW$_\mathrm{th}$ reactor would allow for a more than $5\,\sigma$ detection of the CEvNS signal. Scaling the exposure to 100--1,000\,g-yr would yield  a flux-systematics limited measurement at the percent level. 
\item Reactor monitoring for nuclear security and treaty verification in NaI, CsI, BaF$_2$, BGO and CaWO$_4$: We find that the sensitivity to plutonium production conforms to the IAEA detection goals~\cite{IAEA2001}, both in terms of timeliness and plutonium quantity, for detector masses as small as 100\,g. Given that our system is passive and small, its concept of operation likely fits well within established IAEA workflows and could be an attractive replacement of the thermohydraulic power monitor used at some research reactors. Immunity from cyber-threats, combined with a generally un-spoofable signal, is another attractive feature for these applications.
\end{enumerate}

For all three examples studied, the detectors would be small (10--1,000\,g), passive, robust, cheap and maintenance-free. They would operate unsupervised at room temperature with no electronics. Potentially  only modest passive shielding might be required.

Furthermore, we note that sub-gram detectors would make excellent high-sensitivity, passive neutron detectors for a wide range of neutron energies. These could be especially well suited for zero-knowledge proof protocols for disarmament verification of nuclear warheads~\cite{Glaser2014,Philippe2016}. The idea is that a neutron beam interrogates the warhead, but a pre-loaded neutron image, only known to the owner of the warhead, is superimposed inside the detector. This allows to verify that several objects are the same without revealing any details about their construction. This approach relies on passive neutron detectors which can be reliably pre-loaded by the host, can store the neutron information for a long time and can be read out repeatedly. It appears likely that the detectors proposed here have the requisite combination of properties. 

The quality of the simulation performed here and the uncertainty on inputs, like the threshold damage energy, indicate that a follow-up with experimental studies is needed. We consider this as the best way forward. For a proof of concept, next steps would include investigating to what extent color center formation is selective against gamma rays, which, out of the materials studied here has been demonstrated only for LiF and BaF$_2$~\cite{Mosbacher:2019igk}. An at-scale demonstration of single color center imaging, with the requisite resolution and speed, would also be an important next step. Following on would be detailed studies of neutron recoils and the relationship between neutron energy and color center formation rate, essentially mapping the threshold function. These data could also be used to create validated simulation tools. In combination, these steps would likely allow for an attempt to deploy a detector at a reactor and obtain a first observation of reactor CEvNS in the near future. Since the proposed detector materials and readout systems are commercially available, the necessary R\&D program could be relatively fast and cheap to put together.\\

\section*{Acknowledgements}

This work was supported by the National Science Foundation REU grant
number 1757087, by the U.S. Department
of Energy Office of Science under award number DE-SC00018327 and by the 
National Nuclear Security Administration Office of Defense Nuclear
Nonproliferation R\&D through the Consortium for Monitoring,
Technology and Verification under award number DE-NA0003920. PH would like to acknowledge useful conversations with S.~Baum, A.~Hecht and I.~Jovanovic. BKC would like to acknowledge useful conversations with R.~Goldston.

\bibliography{apssamp}

\clearpage

\begin{appendix}

\section*{Photon economics, heating and data rates}

To scan a given volume per day, $V$, several factors need to be considered
\begin{enumerate}
    \item Can a microscope system cope with the mechanical needs to scan the volume $V$ at the required rate by re-positioning the sample at the requisite precision and focus fast enough?
    \item What is the density of pre-exisiting color centers?
    \item What laser power is need to achieve clear detection of color centers in the volume $V$?
    \item Is the resulting heat input into the sample manageable?
    \item Is the resulting data rate manageable?
\end{enumerate}

\subsection{Mechanics}
To give a specific example, consider spherical-aberration-assisted extended depth-of-field (SPED) light-sheet microscopy as an example~\cite{Tomer2015}, we find that imaging of 12 volumes of $0.9\times0.4\times0.2\,\mathrm{mm}^3$ per second has been demonstrated, which translates to 75$\,\mathrm{cm}^3$ per day. This demonstrates that the mechanics of scanning a large volume is a solved problem.

\subsection{Pre-exisiting color centers}

The results of Ref.~\cite{Mosbacher:2019igk} allow us to put an upper bound on the intrinsic color center density in the materials of interest by estimating the number of radiation induced color centers: Given the neutron flux of $(3.8-8)\times 10^4\,\mathrm{cm}^{-2}\,\mathrm{s}^{-1}$ and a neutron scattering cross section between 2-4 barns and an irradiation time of 64.3\,h, we obtain $10^8-10^9$ neutron interactions per cm$^3$. Based on a TRIM calculation this results in about 200--500 vacancies per neutron event with a neutron energy spectrum corresponding to $^{252}$Cf, or about $10^{10}-10^{12}$ color centers per cm$^3$. Given target atom densities of few times $10^{22}\,\mathrm{cm}^{-3}$, this corresponds to relative radiation induced color center abundances of $f_d<10^{-12}-10^{-10}$. This number presents a conservative upper bound on the intrinsic color center abundance.

\subsection{Power}
For a material of given molecular weight $M$ and density $\rho$, there are 
\begin{equation}
    N=\rho / M N_A\,,
\end{equation}
target sites per cm$^3$, where $N_A$ is Avogadro's constant. Assuming a pre-exisiting defect fraction $f_d$, the density $N_p$ of pre-existing defects is  $N_P=f_d N$. Defining the volume of a voxel $V_\mathrm{vox}$ as
\begin{equation}
    V_\mathrm{vox}= v_x v_y v_z\,,
\end{equation}
where $v_x, v_y, v_z$ are the corresponding voxel dimensions in the $x, y$, and $z$ directions, the number of defects per voxel $D$ is given by $D=N_d V_\mathrm{vox}$.

The number of photons, $p$  emitted by a single color center per unit time is
\begin{equation}
    p=\phi \sigma_0 \xi\,,
\end{equation}
where $\phi$ is the photon fluence into the sample, $\sigma_0=10^{-17}\,\mathrm{cm}^2$ and $\xi$ is the oscillator strength. The number of detected photons per unit time $p_d$ is given as $p_d=p\, \mathrm{NA}^2/4\, \epsilon$, where NA is the numerical aperture of the collection microscope objective and $\epsilon$ is the quantum efficiency of the camera.

Next we need to determine how many photons we need to detect, $p_n$, to be certain that there is one additional color center in a voxel relative to the number of pre-existing color centers $D$ in that voxel. Assuming that the number of photons $\gamma$ is reasonably large, say bigger than 10, Poissonian count rate statistics is well approximated by a normal distribution with mean $\gamma$ and standard deviation $\sqrt{\gamma}$. If we require that out of all voxels at most one spurious detection occurs on average, it follows
\begin{equation}
    \int_{-\infty}^{x}\,d x' \frac{1}{2\pi} e^{-x'^2} = 1-\frac{V_\mathrm{vox}}{V}\,,
\end{equation}
where $x$ sets the number of standard deviations needed. If we have $D$ pre-existing color centers and each color center creates $p_1$ photons then  the number of photons from one additional color center needs to exceed the standard deviation of $p_1D$ photons from the pre-exisiting color centers by a factor $x$:
\begin{equation}
\label{eq:sigs}
    \frac{p_1}{\sqrt{p_1D}}=x \quad\mathrm{or}\quad p_1=2D\,\mathrm{erfc}^{-1}\left(2-2\frac{V_\mathrm{vox}}{V}\right)^2\,,
\end{equation}
where $\mathrm{erfc}^{-1}$ is the inverse complementary error function. Note, that $\mathrm{erfc}^{-1}$  has only a logarithmic dependence on  $V_\mathrm{vox}/V$. To obtain the number of photons which are needed to excite the fluorescence at rate $p_1$ we need to divide $p_1$ by the area $A$ of the beam/light sheet, which is given by $A=v_z\sqrt{\mathrm{fov}}$, where $\mathrm{fov}$ is the area which is imaged at one time, assuming that $\mathrm{fov}$ is quadratic. We then require $p_1/A=p$ and can solve for fluence $\phi$. Note, that not each voxel needs to receive the fluence $\phi$ separately; since the sample is still optically thin, {\it i.e.} only a small fraction of fluorescence photons is absorbed in each voxel. 

Finally, we need to consider time: the number of separate images taken of the volume $V$ is $V/(\mathrm{fov}v_z)$ and thus the time $t$ per image is the $t=T(\mathrm{fov}v_z)/V $, where $T$ is the time alloted to scan all of the volume $T$. The remaining step is to multiply the result with the energy of each photon $h c/\lambda$. We obtain for the beam power of the exciting laser for a total volume of $1\,\mathrm{cm}^3$ and for $T=1$\,day:
\begin{widetext}
\begin{equation}
\label{eq:power}
    P=0.4\,\mathrm{W}\,\left(\frac{V_\mathrm{vox}}{\mu\mathrm{m}^3}\right)\left(\frac{10^6\,\mu\mathrm{m}^2}{\mathrm{fov}}\right)^{1/2}\left(\frac{f_d}{10^{-10}}\right)\left(\frac{\rho}{10\,\mathrm{g}\,\mathrm{cm}^{-3}}\right)
\left(\frac{1}{\xi}\right)\left(\frac{150\,\mathrm{g}\,\mathrm{mol}^{-1}}{M}\right)\left(\frac{250\,\mathrm{nm}}{\lambda}\right) \left(\frac{1.5}{\mathrm{NA}}\right)^2\left(\frac{0.8}{\epsilon}\right) f\left(\frac{V_\mathrm{vox}}{V}\right)\,,
\end{equation}
\end{widetext}
 where $P\propto V/T$ and $f(x)\simeq -0.0954 - 0.0912 \log_{10} x$, where the latter arises from expanding Eq.~\ref{eq:sigs} around $V_\mathrm{vox}/V=10^{-12}$.
 
 \subsection*{Heat}
 
 The laser beam has a cross section of $\sqrt{\mathrm{fov}}v_z$ and per unit depth into the sample deposits a fraction of the beam power proportional to the absorption coefficient $\alpha$, {\it i.e.} $P\alpha$, which results into uniform planar heat input, again assuming that sample is optically thin. Heat conduction then yields a temperature difference perpendicular to the beam of $\Delta T = P\alpha/\lambda \,d$, where $d$ is the thickness of the sample; here we assume that $v_z\ll d$, so that we effectively have a one-dimensional heat conduction problem. In appropriate units this becomes:
 \begin{equation}
 \label{eq:heat}
     \Delta T = 1\,\mathrm{K}\left(\frac{P}{\mathrm{W}}\right)\left(\frac{10\,\mathrm{W}\,\mathrm{K}^{-1}\,\mathrm{m}^{-1}}{\lambda}\right)\left(\frac{\alpha}{0.1\mathrm{cm}^{-1}}\right)\left(\frac{d}{\mathrm{cm}}\right)
 \end{equation}
 The amount of heating is small because our samples are essentially transparent and reasonable heat conductors.

\subsection*{Data}

The data volume $B$ is is given as number of pixels per image $\mathrm{fov}/v_x/v_y$ times the number of images taken $V/(\mathrm{fov}v_z)$
\begin{equation}
\label{eq:data}
    B=2\,\mathrm{TByte}\left(\frac{V}{1\,\mathrm{cm}^2}\right)\left(\frac{\mu\mathrm{m}^3}{V_\mathrm{vox}}\right)\left(\frac{b_d}{16\,\mathrm{bit}}\right)\,,
\end{equation}
 where $b_d$ is the bit depth of the image.

 \subsection*{Discussion}
 
 The required mechanical performance has been demonstrated by existing light-sheet microscopes. We have derived some simple scaling laws for the required laser power, resulting heating and data volume. In the presence of pre-existing color centers, decreasing the voxel volume is a way to keep the required power limited, demonstrating the advantage of voxelization over total volume fluorescence. The laser power is further kept small by imaging an entire field of view at once, in contrast to confocal microscopy, reducing power by 3-4 orders of magnitude. As a result heating is generally very mild once heat conduction and the very small optical absorption of the samples are accounted for. The resulting data volume is significant but not does not seem to present a major obstacle. Consider a specific example of 100\,g of BaF$_2$ to be scanned in one day with a pre-existing color center concentration of $f_d=10^{-10}$ and an excitation wavelength of 250\,nm~\cite{Mosbacher:2019igk}. BaF$_2$ has a density of $4.88\,\mathrm{g}\,\mathrm{cm}^{-3}$ and a molecular weight of 137.3\,u (see Tab.~\ref{tab:materials}). Using $V_\mathrm{vox}=1\,\mu\mathrm{m}^3$, we obtain P=4.89\,W (Eq.~\ref{eq:power}). The heat conductivity of BaF$_2$  $\lambda=7.1\,\mathrm{W}\,\mathrm{m}^{-1}\,\mathrm{K}^{-1}$ and the absorption coefficient around 250\,nm is 0.2$\,\mathrm{cm}^{-1}$~\cite{baf2}. This results in $\Delta T=13.8\,$K for a 1\,cm thick sample from Eq.~\ref{eq:heat}. Note, that in practice a thinner sample would be preferred from an optical point of view related to the working distance of high-performance microscope objectives. For instance, taking a Nikon CFI Apo LWD Lambda S 20XC WI~\cite{nikon} objective as example: it has a working distance of just above 1\,mm (0.9-0.99\,mm plus 0.11-0.23\,mm allowance for the cover slip) and a numerical aperture of NA=0.95. In this case the power is increased by $(1.5/0.95)^2$ to $12.2\,W$. However, the sample could to only be 1\,mm thick to allow for the whole volume to be scanned and thus, the temperature rise would be $\Delta T=3.4\,$K. A 100\,g 1\,mm thick sample of BaF$_2$ would be approximately $15\times 15\,\mathrm{cm}^2$. Finally, utilizing Eq.~\ref{eq:data} we find the data volume for this example is approximately 41\,TByte. We find similar numbers for the other materials listed in Tab.~\ref{tab:materials}.
 
 In synopsis, none of the parameters used here appear to be extreme or problematic, with the key figure of merit being $f_d/\xi$, i.e. the ratio of pre-existing color center density to the oscillator strength. With BaF$_2$ having a melting point of 1600\,K (see Tab.~\ref{tab:materials}), a $\Delta T\simeq 340\,K$, two orders of magnitude higher than the value found using the above above parameters still seems acceptable, so there is considerable head room in our estimate.
 
\end{appendix}

\end{document}